\documentclass[a4paper]{jpconf}
\usepackage{graphicx}
\begin{document}
\title{Search for double beta decay of $^{116}$Cd with enriched $^{116}$CdWO$_4$ crystal scintillators (Aurora experiment)}

\author{F.~A.~Danevich~$^{1}$, A.~S.~Barabash~$^{2}$, P.~Belli~$^{3,4}$, R.~Bernabei~$^{3,4}$, F.~Cappella~$^{5}$, V.~Caracciolo~$^{5}$, R.~Cerulli~$^{5}$, D.~M.~Chernyak~$^{1}$, S.~d'Angelo~~$^{3,4,}$\footnote[6]{Deceased}, A.~Incicchitti~$^{7,8}$, V.~V.~Kobychev~$^{1}$, S.~I.~Konovalov~$^{2}$, M.~Laubenstein~$^{5}$, V.~M.~Mokina~$^{1,7}$, D.~V.~Poda~$^{1,9}$, O.~G.~Polischuk~$^{1}$, V.~N.~Shlegel~$^{10}$, V.~I.~Tretyak~$^{1,7}$ and V.~I.~Umatov~$^{2}$
}

\address{$^{1}$~Institute for Nuclear Research, MSP 03680 Kyiv, Ukraine

 $^{2}$~National Research Centre ``Kurchatov Institute", Institute of Theoretical and Experimental Physics, 117218 Moscow, Russia

 $^{3}$~INFN, sezione di Roma ``Tor Vergata'', I-00133 Rome, Italy

 $^{4}$~Dipartimento di Fisica, Universit\`a di Roma ``Tor Vergata'', I-00133 Rome, Italy

 $^{5}$~INFN, Laboratori Nazionali del Gran Sasso, I-67100 Assergi (AQ), Italy

 $^{7}$~INFN, sezione di Roma, I-00185 Rome, Italy

 $^{8}$~Dipartimento di Fisica, Universit\`a di Roma ``La Sapienza'', I-00185 Rome, Italy

 $^{9}$~CSNSM, Univ. Paris-Sud, CNRS/IN2P3, Universit\'{e} Paris-Saclay, 91405 Orsay, France
% $^{9}$~Centre de Sciences Nucl\'eaires et de Sciences de la Mati\`ere, 91405 Orsay, France

 $^{10}$~Nikolaev Institute of Inorganic Chemistry, 630090 Novosibirsk, Russia

 }

\ead{danevich@kinr.kiev.ua}

\begin{abstract}

The Aurora experiment to investigate double beta decay of
$^{116}$Cd with the help of 1.162 kg cadmium tungstate crystal
scintillators enriched in $^{116}$Cd to 82\% is in progress at the
Gran Sasso Underground Laboratory. The half-life of $^{116}$Cd
relatively to the two neutrino double beta decay is measured with
the highest up-to-date accuracy
$T_{1/2}=(2.62\pm0.14)\times10^{19}$ yr. The sensitivity of the
experiment to the neutrinoless double beta decay of $^{116}$Cd to
the ground state of $^{116}$Sn is estimated as $T_{1/2} \geq
1.9\times10^{23}$ yr at 90\% CL, which corresponds to the
effective Majorana neutrino mass limit $\langle m_{\nu}\rangle
\leq (1.2-1.8)$ eV. New limits are obtained for the double beta
decay of $^{116}$Cd to the excited levels of $^{116}$Sn, and for
the neutrinoless double beta decay with emission of majorons.

\end{abstract}

\section{Introduction}

Observations of neutrino oscillations give a clear evidence of
effects beyond the Standard Model of particles (see, e.g., review
\cite{Mohapatra:2007}) and provide a strong motivation to
investigate neutrinoless double beta ($0\nu2\beta$) decay of
atomic nuclei. The $0\nu2\beta$ decay violates the lepton-number
conservation and is only possible if neutrino is a massive
Majorana particle. Therefore, search for $0\nu2\beta$ decay is
considered as a promising way to clarify the nature of the
neutrino, check the lepton number conservation, determine the
absolute scale of the neutrino mass and the neutrino mass
hierarchy, test the existence of effects beyond the Standard
Model, in particular, existence of hypothetical Nambu-Goldstone
bosons (majorons) and right-handed currents in weak interaction
\cite{Vergados:2012,Barea:2012,Rodejohann:2012,Deppisch:2012,Bilenky:2015,Pas:2015}.

The isotope $^{116}$Cd is one of the most promising $2\beta$
nuclei thanks to the favorable theoretical estimations of the
decay probability (\cite{Vergados:2012,Barea:2012}), large energy
release $Q_{2\beta} = 2813.50(13)$ keV \cite{Rahaman:2011},
relatively high isotopic abundance $\delta=7.49\%$
\cite{Berglund:2011} and a possibility of isotopic enrichment in a
large amount.

Experimental investigations of $^{116}$Cd $2\beta$ decay were
realized by tracking detectors with enriched cadmium foil as
source \cite{Ejiri:1995,Arnold:1996,Tretyak:2013}, and by
calorimetric approach using CdWO$_4$ crystal scintillators
\cite{Danevich:1995,Danevich:2000,Danevich:2003} and CdZnTe room
temperature semiconductors \cite{Ebert:2013}. The $2\beta$ decay
to excited levels of $^{116}$Sn were also searched for with low
background HPGe $\gamma$ detectors
\cite{Barabash:1990,Piepke:1994}. Large volume radiopure cadmium
tungstate crystal scintillators were produced from cadmium
enriched in $^{116}$Cd to 82\% ($^{116}$CdWO$_4$) to investigate
double beta decay of $^{116}$Cd \cite{Barabash:2011}. The crystals
show excellent scintillation properties and low level of
radioactive contamination \cite{Poda:2013,Danevich:2013}.
Preliminary results of the Aurora experiment were reported in the
conference proceedings \cite{Poda:2014,Polischuk:2016}. Here we
present recent results of the experiment.

\section{Experiment}

Two $^{116}$CdWO$_4$ crystal scintillators with a total mass 1.162
kg ($1.584 \times 10^{24}$ of $^{116}$Cd nuclei) are installed in
the low background DAMA/R\&D set-up operated at the Gran Sasso
Underground Laboratory of I.N.F.N. (Italy). The low background
set-up with the $^{116}$CdWO$_4$ detectors has been modified
several times to improve the energy resolution and to decrease
background. In the last configuration of the set-up the
$^{116}$CdWO$_4$ crystal scintillators are fixed in
polytetrafluoroethylene containers filled with ultrapure liquid
scintillator. The liquid scintillator improves the light
collection from the $^{116}$CdWO$_4$ crystal scintillators and
serves as an anti-coincidence veto counter. The scintillators are
viewed through high purity quartz light-guides ($\oslash 7\times
40$ cm) by low background high quantum efficiency photomultiplier
tubes (PMT, Hamamatsu R6233MOD). The detectors are installed
inside a low radioactive copper box flushed with high pure
nitrogen gas with an external shield made of radiopure materials:
copper (15 cm), lead (15 cm), cadmium (1.5 mm) and paraffin (4 to
10 cm). The whole set-up is enclosed in a plexiglas box also
flushed with high purity nitrogen gas to remove radon.  An
event-by-event data acquisition system based on a 1 GS/s 8 bit
transient digitizer (Acqiris DC270) records time and pulse profile
of events. The energy scale and the energy resolution of the
detector are checked periodically with $^{22}$Na, $^{60}$Co,
$^{137}$Cs, and $^{228}$Th $\gamma$ sources. The energy resolution
of the $^{116}$CdWO$_4$ detector for 2615 keV $\gamma$ quanta of
$^{208}$Tl is FWHM$~\approx5\%$.

\section{Results and discussion}

\begin{figure}[h]
\includegraphics[width=18pc]{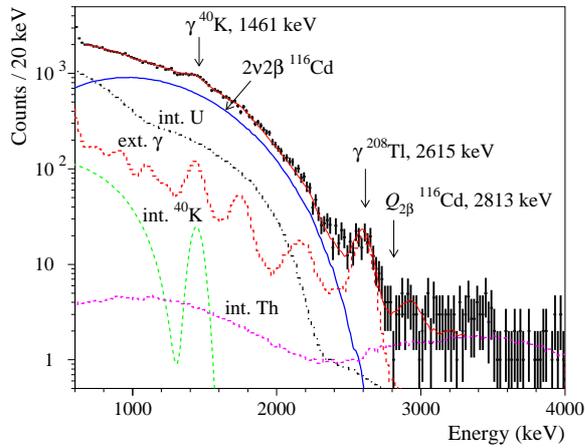}\hspace{2pc}%
\begin{minipage}[b]{18pc}\caption{\label{2n2b}The energy spectrum of $\beta$ and $\gamma$
events accumulated over 12015 h together with the main components
of the background model: $2\nu2\beta$ decay of $^{116}$Cd (``$2\nu
2\beta $ $^{116}$Cd"), the distributions of the internal
contamination of the $^{116}$CdWO$_4$ crystals by potassium
(``int. $^{40}$K"), thorium (``int. Th") and uranium (``int. U"),
and the contribution from external $\gamma$ quanta (``ext.
$\gamma$").}
\end{minipage}
\end{figure}

The energy spectrum of $\beta$ and $\gamma$ events accumulated
over 12015 h by the $^{116}$CdWO$_4$ detectors is presented in
Fig. \ref{2n2b}. The $\beta$ and $\gamma$ events were selected
with the help of two pulse-shape discrimination methods: the
optimal filter method to select $\alpha$ particles, and the front
edge analysis to select Bi--Po events (fast sub-chains
$^{212}$Bi--$^{212}$Po and $^{214}$Bi--$^{214}$Po from $^{232}$Th
and $^{238}$U chains, respectively) from internal contamination of
the crystals by U and Th. Besides, both the pulse-shape
discrimination techniques are also sensitive to pile-ups of
$^{116}$CdWO$_4$ and liquid scintillator signals. The experimental
spectrum was fitted in the energy interval $(660-3300)$ keV by the
model constructed from the two neutrino double beta ($2\nu2\beta$)
spectrum of $^{116}$Cd, the distributions of the $^{116}$CdWO$_4$
crystal scintillators internal contamination by potassium, thorium
and uranium (taking into account possible disequilibrium of the
$^{232}$Th and $^{238}$U chains), and the contribution from
external $\gamma$ quanta (from radioactive contamination of the
PMTs, quartz light-guides and copper of the passive shield).
Response of the $^{116}$CdWO$_4$ detector to the $2\beta$
processes in $^{116}$Cd as well as to the radioactive
contamination of the set-up were simulated with EGS4 package
\cite{EGS4}. The initial kinematics of the particles emitted in
the decay of the nuclei was given by an event generator DECAY0
\cite{DECAY0}. The fit gives the following half-life of $^{116}$Cd
relatively to the $2\nu2\beta$ decay to the ground state of
$^{116}$Sn:

 \begin{center}
$T_{1/2}^{2\nu2\beta} = [2.62 \pm 0.02(stat.) \pm 0.14(syst.)]
\times 10^{19}$ yr.
 \end{center}

\begin{figure}[h]
\includegraphics[width=15pc]{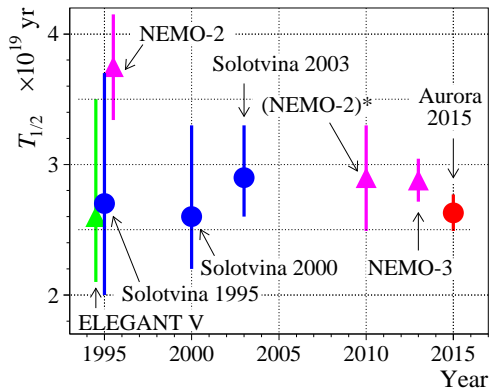}\hspace{2pc}
\begin{minipage}[b]{21pc}\caption{\label{comparison}Comparison of the $^{116}$Cd $2\nu2\beta$
half-life obtained in the Aurora experiment with other
experiments: ELEGANT V \cite{Ejiri:1995}, Solotvina
\cite{Danevich:1995,Danevich:2000,Danevich:2003}, NEMO-2
\cite{Arnold:1996} and NEMO-3 \cite{Tretyak:2013}. A reevaluated
NEMO-2 value \cite{Barabash:2010} is labelled as (NEMO-2)*.}
\end{minipage}
\end{figure}

The main sources of the systematic error are the uncertainties of
the radioactive contamination of the crystal scintillators and of
the details of the set-up, and variation of the effect's area
depending on the interval of the fit. The signal to background
ratio is 2.6:1 in the energy interval $(1.1-2.8)$ MeV. The
comparison of the $^{116}$Cd $2\nu2\beta$ half-life obtained in
the Aurora experiment with other experiments is given in Fig.
\ref{comparison}. The result is in agreement with the previous
experiments
\cite{Ejiri:1995,Arnold:1996,Tretyak:2013,Danevich:1995,Danevich:2000,Danevich:2003},
however the half-life of $^{116}$Cd is determined in the present
study with the highest accuracy.

\begin{figure}[h]
\includegraphics[width=17pc]{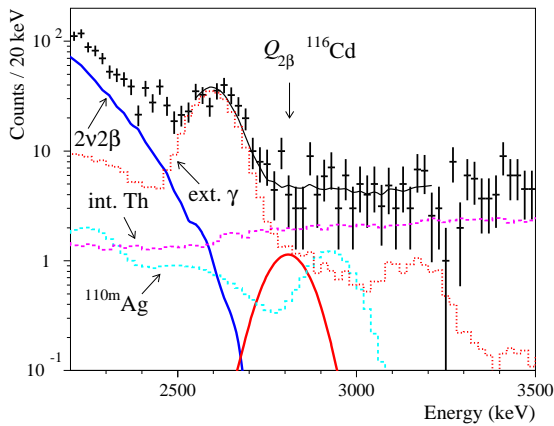}\hspace{2pc}
\begin{minipage}[b]{18pc}\caption{\label{0n2b}The energy spectrum of $\beta$ and $\gamma$
events accumulated over 20711 h with the $^{116}$CdWO$_4$
detectors in the region of interest together with the background
model: the $2\nu2\beta$ decay of $^{116}$Cd (``$2\nu 2\beta $"),
the internal contamination of the $^{116}$CdWO$_4$ crystals by
cosmogenic $^{110m}$Ag (``$^{110m}$Ag") and $^{228}$Th (``int.
Th"), and the contribution from external $\gamma$ quanta (``ext.
$\gamma$").}
\end{minipage}
\end{figure}

There are no other peculiarities in the experimental data which
could be interpreted as $2\beta$ processes in $^{116}$Cd. To
estimate limit on $0\nu2\beta$ decay of $^{116}$Cd to the ground
state of $^{116}$Sn we have used data of two runs with the lowest
background in the region of interest: the current one and the
accumulated over 8696 h in the set-up described in
\cite{Poda:2014}. The sum energy spectrum is presented in Fig.
\ref{0n2b}. The background counting rate of the detector in the
energy interval $(2.7-2.9)$ MeV (which contains 80\% of the
$0\nu2\beta$ distribution) is $\approx0.1$
counts/(yr$\times$kg$\times$keV). A fit of the spectrum in the
energy interval $(2560-3200)$ keV by the background model
constructed from the distributions of the $0\nu2\beta$ decay of
$^{116}$Cd (effect searched for), the $2\nu2\beta$ decay of
$^{116}$Cd with the half-life $2.62\times10^{19}$ yr, the internal
contamination of the crystals by $^{110m}$Ag and $^{228}$Th, and
the contribution from external $\gamma$ quanta gives an area of
the expected peak $S=-3.7\pm10.2$, which gives no evidence of the
effect. In accordance with  \cite{Feldman:1998} 13.3 counts are
excluded at 90\% confidence level. Taking into account the 99\%
efficiency of the pulse-shape discrimination to select $\beta$
($\gamma$) events and 99\% efficiency of the front edge analysis
(98\% in total), we got the following new limit on the
$0\nu2\beta$ decay of $^{116}$Cd to the ground state of
$^{116}$Sn:

 \begin{center}
$T_{1/2}^{0\nu2\beta} \geq 1.9 \times 10^{23}$ yr.
 \end{center}

The half-life limit corresponds to the effective neutrino mass
limit $\langle m_{\nu}\rangle \leq (1.2-1.8)$ eV, obtained by
using the recent nuclear matrix elements reported in
\cite{Rodriguez:2010,Simkovic:2013,Hyvarinen:2015,Barea:2015}, the
phase space factor from \cite{Kotila:2012} and the value of the
axial vector coupling constant $g_{A}=1.27$.

Limits on $2\beta$ decay processes in $^{116}$Cd to the excited
levels of $^{116}$Sn, and for the $0\nu2\beta$ decay with emission
of one ($\chi$), two ($2\chi$) and bulk ($\chi^{bulk}$) majorons
were derived from the fits of the data in the energy intervals
with a high effect to background ratio. The results are presented
in Table \ref{results}. Using the bound on the $0\nu2\beta$ decay
with one majoron emission and the same calculations of the nuclear
matrix elements we have estimated a limit on the effective majoron
neutrino coupling constant $g_{\nu\chi}\leq(5.3-8.5)\times
10^{-5}$.

\begin{table}[h]
\caption{\label{results} The half-life limits and half-life value
on the $2\beta$ decay processes in $^{116}$Cd. The most stringent
limits obtained in the previous experiments are given for
comparison. The limits are given at the 90\% CL except the results
\cite{Barabash:1990}, which are given at the 68\% CL.}
\begin{center}
\begin{tabular}{llll}
\br
 Decay mode     & Transition,           & $T_{1/2}$ (yr)            &  Previous \\
 ~              & level of $^{116}$Cd   & ~                         &  result \\
 ~              & (keV)                 & ~                         & \\
 \mr

 $0\nu$         & g.s.                  & $\geq 1.9\times10^{23}$  & $\geq 1.7\times10^{23}$ \cite{Danevich:2003} \\
 $0\nu$         & $2^+(1294)$           & $\geq 6.2\times10^{22}$  & $\geq 2.9\times10^{22}$ \cite{Danevich:2003} \\
 $0\nu$         & $0^+(1757)$           & $\geq 6.3\times10^{22}$  & $\geq 1.4\times10^{22}$ \cite{Danevich:2003} \\
 $0\nu$         & $0^+(2027)$           & $\geq 4.5\times10^{22}$  & $\geq 6.0\times10^{21}$ \cite{Danevich:2003} \\
 $0\nu$         & $2^+(2112)$           & $\geq 3.6\times10^{22}$  & $\geq 1.7\times10^{20}$ \cite{Barabash:1990} \\
 $0\nu$         & $2^+(2225)$           & $\geq 4.1\times10^{22}$  & $\geq 1.0\times10^{20}$ \cite{Barabash:1990} \\
 $0\nu \chi$    & g.s.                  & $\geq 1.1\times10^{22}$  & $\geq 8.0\times10^{21}$ \cite{Danevich:2003} \\
 $0\nu 2\chi$   & g.s.                  & $\geq 9.0\times10^{20}$  & $\geq 8.0\times10^{20}$ \cite{Danevich:2003} \\
 $0\nu \chi^{bulk}$   & g.s.            & $\geq 2.1\times10^{21}$  & $\geq 1.7\times10^{21}$ \cite{Danevich:2003} \\
 $2\nu$         & g.s.                  & $=(2.62\pm0.14)\times10^{19}$     & see Fig. \ref{comparison} \\
 $2\nu$         & $2^+(1294)$           & $\geq 9.0\times10^{20}$  & $\geq 2.3\times10^{21}$ \cite{Piepke:1994} \\
 $2\nu$         & $2^+(1757)$           & $\geq 1.0\times10^{21}$  & $\geq 2.0\times10^{21}$ \cite{Piepke:1994} \\
 $2\nu$         & $2^+(2027)$           & $\geq 1.1\times10^{21}$  & $\geq 2.0\times10^{21}$ \cite{Piepke:1994} \\
 $2\nu$         & $2^+(2112)$           & $\geq 2.3\times10^{21}$  & $\geq 1.7\times10^{20}$ \cite{Barabash:1990} \\
 $2\nu$         & $2^+(2225)$           & $\geq 2.5\times10^{21}$  & $\geq 1.0\times10^{20}$ \cite{Barabash:1990} \\

\br
\end{tabular}
\end{center}
\end{table}

\section{Conclusions}

The Aurora experiment is in progress to investigate $2\beta$
processes in $^{116}$Cd by using enriched $^{116}$CdWO$_4$
scintillation detectors. The $2\nu2\beta$ half-life of $^{116}$Cd
is measured with the highest up-to-date accuracy: $T_{1/2} = (2.62
\pm 0.14)\times 10^{19}$ yr. The new improved $0\nu 2\beta$
half-life limit was set as $T_{1/2}\geq 1.9\times10^{23}$ yr at
90\% CL, which corresponds to the effective Majorana neutrino mass
$\langle m_{\nu}\rangle \leq (1.2-1.8)$ eV. New limits on the
$2\beta$ decay to excited levels of $^{116}$Sn and the
$0\nu2\beta$ decay with emission of one, two and bulk majorons
were set at the level of $T_{1/2} \geq (10^{20}-10^{22})$ yr.
Using the limit $T_{1/2}\geq 1.1\times10^{22}$ yr on the
$0\nu2\beta$ decay with one majoron emission we have obtained one
of the strongest limits on the effective majoron neutrino coupling
constant $g_{\nu\chi}\leq(5.3-8.5)\times 10^{-5}$. It is worth
noting that we have observed a segregation of thorium, radium and
potassium in the crystal growing process, which provides a
possibility to improve substantially the radiopurity of the
$^{116}$CdWO$_4$ crystal scintillators by re-crystallization,
which is in progress now.

\end{document}